\documentclass[a4paper]{article}

\usepackage[margin=1in]{geometry}

\usepackage[T1]{fontenc}
\newcommand{\changefont}[3]{
\fontfamily{#1} \fontseries{#2} \fontshape{#3} \selectfont}

\changefont{ptm}{m}{n}

\usepackage{setspace} 

\usepackage{graphicx}

\usepackage{amsfonts}
\usepackage{amssymb}
\usepackage{graphics}
\usepackage{mathrsfs}
 
\usepackage{color}

\long\def\symbolfootnote[#1]#2{\begingroup%
\def\thefootnote{\fnsymbol{footnote}}\footnote[#1]{#2}\endgroup} 

\begin{document}

\begin{center}
\Large \textbf{Multiple Chaotic Attractors in Coupled Lorenz Systems}
\end{center}

\begin{center}
\normalsize \textbf{Mehmet Onur Fen$^{a,}\symbolfootnote[1]{Corresponding Author Tel.: +90 312 585 0217, E-mail: monur.fen@gmail.com}$} \\
\vspace{0.2cm}
\textit{\textbf{\footnotesize$^a$Department of Mathematics, TED University, 06420 Ankara, Turkey}} \\
\vspace{0.1cm}
\end{center}

\vspace{0.3cm}

\begin{center}
\textbf{Abstract}
\end{center}

\noindent\ignorespaces
Unidirectionally coupled Lorenz systems in which the drive possesses a chaotic attractor and the response admits two stable equilibria in the absence of the driving is under investigation. It is found that double chaotic attractors coexist in the dynamics. The approach is applicable for chains of coupled Lorenz systems. The existence of four chaotic attractors in three coupled Lorenz systems is also demonstrated. 

\vspace{0.2cm}
 
\noindent\ignorespaces \textbf{Keywords:}  Multiple chaotic attractors; Lorenz system; Unidirectional coupling

\vspace{0.6cm}

\section{Introduction} \label{sec1}

The system of differential equations
\begin{eqnarray}
\begin{array}{l} \label{lorenz1}
\dot{x}_1 = -\sigma x_1 + \sigma x_2 \\
\dot{x}_2 = - x_1x_3 +r x_1 - x_2\\
\dot{x}_3 = x_1 x_2 - b x_3,
\end{array}
\end{eqnarray}
where $\sigma$, $r$, and $b$ are constants, was presented by Lorenz \cite{Lorenz63} to investigate the dynamics of the atmosphere. System (\ref{lorenz1}) is capable of exhibiting stable equilibria, periodic orbits, homoclinic explosions, period-doubling bifurcations, and chaos with different values of $\sigma$, $r$, and $b$ \cite{Sparrow82}.

Extension of chaos in dynamics of unidirectionally coupled Lorenz systems in which the response admits either a stable equilibrium point or a stable limit cycle in the absence of driving was demonstrated in study \cite{Akh15}. Moreover, it was shown in paper \cite{Fen17} that under certain conditions chaos is present in the dynamics of the response system even if generalized synchronization is not present. 
Motivated by the results of \cite{Akh15} and \cite{Fen17}, in this paper we also investigate unidirectionally coupled Lorenz systems, but this time the response is system with two stable equilibrium points in the absence of driving. The main purpose of the present study is the demonstration of the coexistence of multiple chaotic attractors in the dynamics of Lorenz systems when they are coupled in a unidirectional way. The coexistence of two and four chaotic attractors are shown.

The paper is structured as follows. In Section \ref{sec2} the model of coupled Lorenz systems is introduced and the presence of sensitivity, which is the main ingredient of chaos \cite{Lorenz63,Wiggins88}, is discussed from the theoretical point of view. The coexistence of double chaotic attractors in the dynamics of unidirectionally coupled Lorenz systems is demonstrated in Section \ref{sec3}. Section \ref{sec4}, on the other hand, is devoted to the coexistence of four chaotic attractors in three coupled Lorenz systems. Finally, concluding remarks are provided in Section \ref{sec5}.

\section{The model} \label{sec2}

We consider the drive system in the form of (\ref{lorenz1}) and the response system is
\begin{eqnarray}
\begin{array}{l} \label{lorenz2}
\dot{y}_1 = - \overline{\sigma} y_1 + \overline{\sigma} y_2 + \mu_1 f_1(x(t)) \\
\dot{y}_2 = -y_1 y_3 +\overline{r} y_1 -y_2 + \mu_2 f_2(x(t)) \\
\dot{y}_3 = y_1 y_2-\overline{b} y_3 + \mu_3 f_3(x(t)),
\end{array}
\end{eqnarray}
where $\overline{\sigma},$ $\overline{r}$, $\overline{b}$, $\mu_1$, $\mu_2$, $\mu_3$ are constants, $x(t)=(x_1(t), x_2(t), x_3(t))$ is a solution of (\ref{lorenz1}), and the functions $f_1$, $f_2$, and $f_3$ are continuous. We mainly assume that the drive system (\ref{lorenz1}) is chaotic, i.e., it admits sensitivity and infinitely many unstable periodic orbits embedded in the chaotic attractor, and that the constants $\overline{\sigma}$, $\overline{r}$, $\overline{b}$ in the response system (\ref{lorenz2}) are such that the Lorenz system
\begin{eqnarray}
\begin{array}{l} \label{nonperturbed_lorenz_system}
\dot{u}_1 = - \overline{\sigma} u_1 + \overline{\sigma} u_2 \\
\dot{u}_2 = - u_1u_3 +\overline{r} u_1 -u_2  \\
\dot{u}_3 = u_1u_2-\overline{b}u_3,
\end{array}
\end{eqnarray}
admits two stable equilibrium points. 

It is worth noting that the coupled system (\ref{lorenz1})-(\ref{lorenz2}) has a skew product structure. Since the drive (\ref{lorenz1}) is chaotic, it possesses a compact invariant set $\Lambda \subset \mathbb R^3$.
Another assumption on system (\ref{lorenz2}) is the existence of a positive number $L$ satisfying $$\left\|f(x) - f(\overline{x})\right\| \ge L \left\|x-\overline{x}\right\|$$ for all $x,$ $\overline{x} \in \Lambda$, where the function $f:\Lambda \to \mathbb R^3$ is defined by $f(x)=(\mu_1 f_1(x),\mu_2 f_2(x),\mu_3 f_3(x))$ and $\left\|.\right\|$ denotes the usual Euclidean norm in $\mathbb R^3$.

We call a solution $x(t)$ of the drive (\ref{lorenz1}) satisfying $x(0)=x_0$, where $x_0$ is a point which belongs to the chaotic attractor of the system, a chaotic solution since it is used as a perturbation in (\ref{lorenz2}). Chaotic solutions may be irregular as well as regular, i.e., periodic and unstable \cite{Lorenz63,Sparrow82,Wiggins88}.

System (\ref{lorenz1}) is called sensitive if there exist positive numbers $\epsilon_0$ and $\Delta$ such that for an arbitrary positive number $\delta_0$ and for each chaotic solution $x(t)$ of (\ref{lorenz1}), there exist a chaotic solution $\overline{x}(t)$ of the system and an interval $J \subset [0,\infty)$ with a length no less than $\Delta$ such that $\left\|x(0)-\overline{x}(0)\right\|<\delta_0$ and $\left\|x(t)-\overline{x}(t)\right\| > \epsilon_0$ for every $t$ in $J$ \cite{Akh15,Fen17}.  

For the discussion of sensitivity in the response system (\ref{lorenz2}), we require that the system possesses a compact invariant set $\mathscr{U} \subset \mathbb R^3$ for each chaotic solution $x(t)$ of (\ref{lorenz1}). The existence of such an invariant set can be shown, for instance, by means of Lyapunov functions \cite{Akh15,Yoshizawa75}.

Let us denote by $\phi_{x(t)}(t,y_0)$ the solution of (\ref{lorenz2}) satisfying $\phi_{x(t)}(0,y_0)=y_0$, where $x(t)$ is a solution of the drive (\ref{lorenz1}) and $y_0$ is a point in $\mathscr{U}$.
We say that system (\ref{lorenz2}) is sensitive if there exist positive numbers $\epsilon_1$ and $\widetilde{\Delta}$ such that for an arbitrary positive number $\delta_1,$ each $y_0\in \mathscr{U}$, and a chaotic solution $x(t)$ of (\ref{lorenz1}), there exist $y_1\in \mathscr{U}$, a chaotic solution $\overline{x}(t)$ of (\ref{lorenz1}), and an interval $\widetilde J \subset [0,\infty)$ with a length no less than $\widetilde{\Delta}$ such that $\left\|y_0-y_1\right\|<\delta_1$ and $\left\|\phi_{x(t)}(t,y_0)-\phi_{\overline{x}(t)}(t,y_1)\right\| > \epsilon_1$ for all $t$ in $\widetilde J$ \cite{Akh15,Fen17}.

It can be proved in a very similar way to Theorem 3.1 presented in paper \cite{Fen17} that the response system (\ref{lorenz2}) is sensitive.

The coexistence of two chaotic attractors in the dynamics of the coupled system (\ref{lorenz1})-(\ref{lorenz2}) is demonstrated in the next section, provided that the constants $\mu_1$, $\mu_2$, and $\mu_3$ are sufficiently small in absolute value.

\section{Coexistence of two chaotic attractors} \label{sec3}

In order to demonstrate the coexistence of two chaotic attractors in the dynamics of the $6$-dimensional system (\ref{lorenz1})-(\ref{lorenz2}),  we set $\sigma=10$, $r=28$, and $b=8/3$ such that the drive system (\ref{lorenz1}) admits a chaotic attractor \cite{Lorenz63,Sparrow82}. Additionally, we consider the response system (\ref{lorenz2}) with $\overline \sigma=10$, $\overline r=12$, $\overline b=8/3$, $\mu_1=\mu_2=\mu_3=0.1$, and $f_1(x_1,x_2,x_3)=7.3x_1+\cos x_1$, $f_2(x_1,x_2,x_3)=1.2 x_2+0.5 \arctan x_2$, $f_3(x_1,x_2,x_3)= 3.5 x_3+0.9 e^{-x_3}$. The points $(-2\sqrt{22/3},-2\sqrt{22/3},11)$ and $(2\sqrt{22/3},2\sqrt{22/3},11)$ are the stable equilibria of system (\ref{nonperturbed_lorenz_system}) with the aforementioned choices of the parameters $\overline \sigma$, $\overline r$, and $\overline b$ \cite{Sparrow82}.

\begin{figure}[ht!] 
\centering
\includegraphics[width=10.0cm]{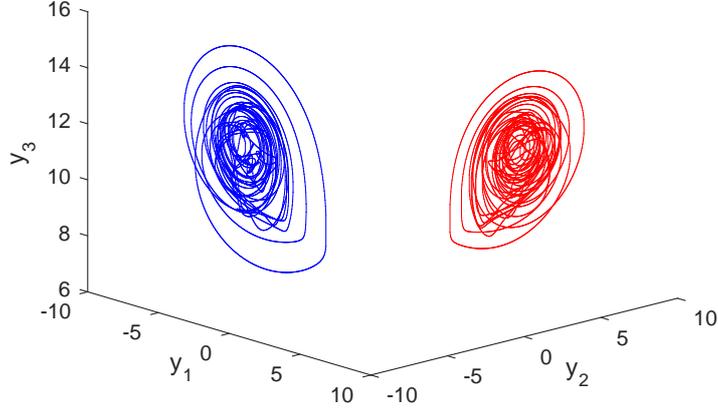}
\caption{Chaotic trajectories of the $6$-dimensional system (\ref{lorenz1})-(\ref{lorenz2}). The figure reveals the coexistence of two chaotic attractors.}
\label{fig1}
\end{figure}

\begin{figure}[ht!] 
\centering
\includegraphics[width=16.0cm]{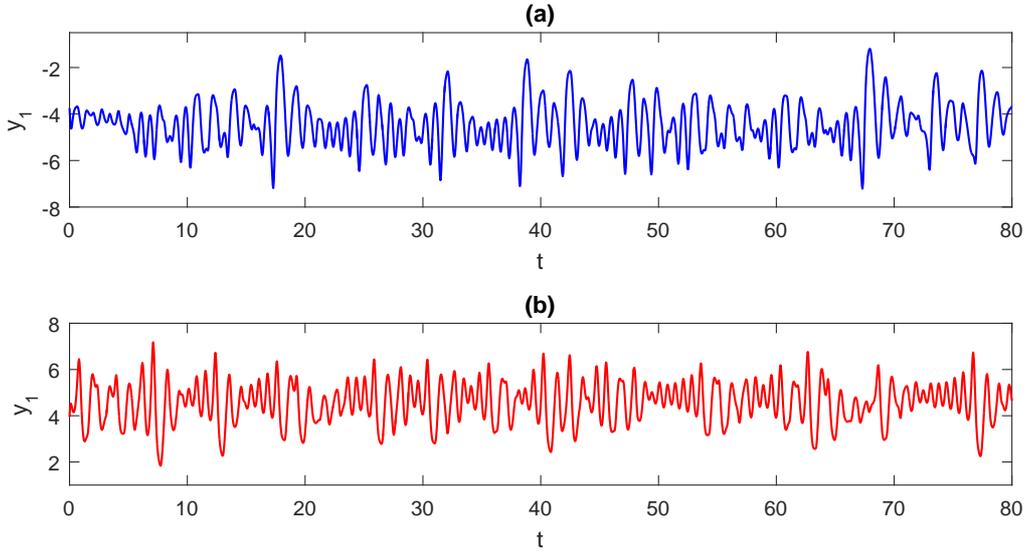}
\caption{Time series of the $y_1$ coordinates of two chaotic solutions of the coupled system (\ref{lorenz1})-(\ref{lorenz2}). (a) The time series corresponding to the initial data $x_1(0)=-15.482$, $x_2(0)=-23.989$, $x_3(0)=26.492$, $y_1(0)=-3.747$, $y_2(0)=-3.493$, $y_3(0)=10.678$; (b) The time series corresponding to the initial data $x_1(0)=13.561$, $x_2(0)=10.162$, $x_3(0)=36.951$, $y_1(0)=3.948$, $y_2(0)=4.187$, $y_3(0)=11.531$.}
\label{fig2}
\end{figure}

Figure \ref{fig1} depicts the projections of two chaotic trajectories of the coupled system (\ref{lorenz1})-(\ref{lorenz2}) on the $y_1 y_2 y_3$-space. The trajectory in blue corresponds to the initial data $x_1(0)=-15.482$, $x_2(0)=-23.989$, $x_3(0)=26.492$, $y_1(0)=-3.747$, $y_2(0)=-3.493$, $y_3(0)=10.678$, whereas the trajectory in red corresponds to the initial data $x_1(0)=13.561$, $x_2(0)=10.162$, $x_3(0)=36.951$, $y_1(0)=3.948$, $y_2(0)=4.187$, $y_3(0)=11.531$. Figure \ref{fig2}, on the other hand, shows the time series of the $y_3$-coordinates of these trajectories. Figure \ref{fig2} (a) and (b) respectively represent the time series of the trajectory in blue and the trajectory in red shown in Figure \ref{fig1}. Both Figure \ref{fig1} and Figure \ref{fig2} manifest that double chaotic attractors coexist in the dynamics of the coupled system (\ref{lorenz1})-(\ref{lorenz2}). 

\section{Coexistence of four chaotic attractors} \label{sec4} 
In this section we will show the coexistence of four chaotic attractors in three coupled Lorenz systems. For that purpose, in addition to the coupled system (\ref{lorenz1})-(\ref{lorenz2}), we set up the system 
\begin{eqnarray}
\begin{array}{l} \label{lorenz3}
\dot{z}_1 = - 10 z_1 + 10 z_2 + 0.19 y_1(t) \\
\dot{z}_2 = -z_1 z_3 +3.6 z_1 -z_2 + 0.23 y_2(t) \\
\dot{z}_3 = z_1 z_2- \displaystyle \frac{8}{3} z_3 + 0.14 y_3(t),
\end{array}
\end{eqnarray} 
where $y(t)=(y_1(t),y_2(t),y_3(t))$ is a solution of (\ref{lorenz2}). Considering the coupling between the systems (\ref{lorenz2}) and (\ref{lorenz3}), system (\ref{lorenz2}) is the drive and system (\ref{lorenz3}) is the response. The parameters of system (\ref{lorenz3}) are such that the system possesses two stable equilibrium points in the absence of the driving, that is, the Lorenz system 
\begin{eqnarray}
\begin{array}{l} \label{lorenz4}
\dot{v}_1 = - 10 v_1 + 10 v_2 \\
\dot{v}_2 = - v_1 v_3 + 3.6 v_1 -v_2  \\
\dot{v}_3 = v_1 v_2- \displaystyle \frac{8}{3} v_3
\end{array}
\end{eqnarray}
admits the stable equilibrium points $(-\sqrt{104/15},-\sqrt{104/15},13/5)$ and $(-\sqrt{104/15},-\sqrt{104/15},13/5)$ \cite{Sparrow82}.

\begin{figure}[ht!] 
\centering
\includegraphics[width=13.0cm]{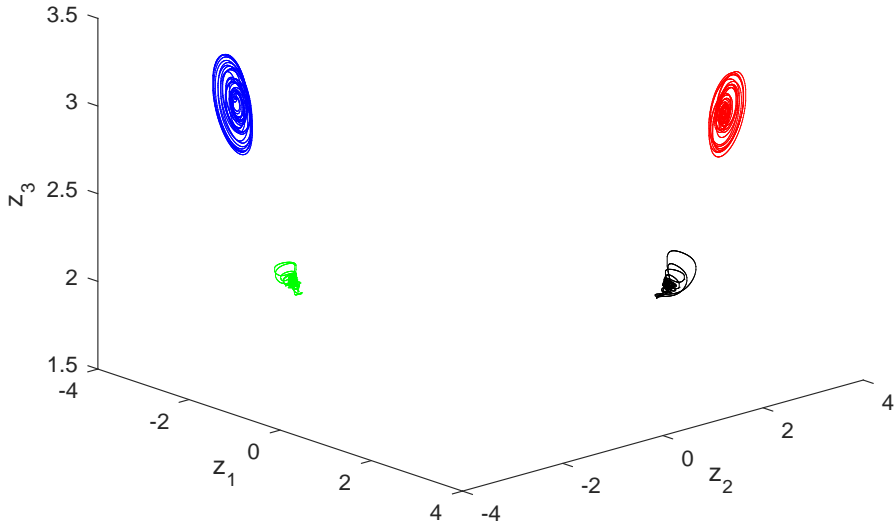}
\caption{Projections of four chaotic trajectories of the $9$-dimensional coupled Lorenz systems (\ref{lorenz1})-(\ref{lorenz2})-(\ref{lorenz3}) on the $z_1z_2z_3$-space. The initial data of the trajectories are provided in Table \ref{table1}. The figure confirms the coexistence of four chaotic attractors in the dynamics.}
\label{fig3}
\end{figure}

\begin{figure}[ht!] 
\centering
\includegraphics[width=13.0cm]{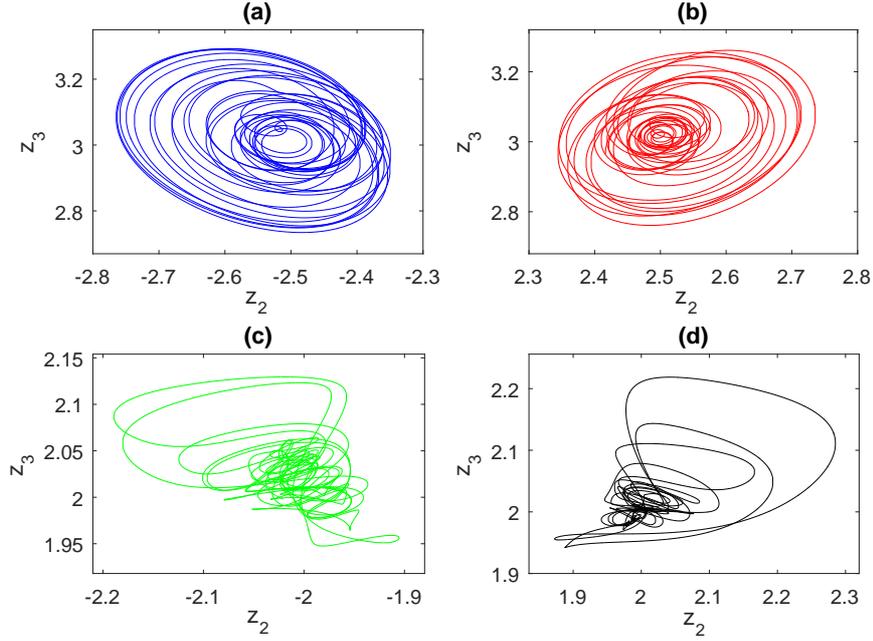}
\caption{Projections of four chaotic trajectories of the $9$-dimensional system (\ref{lorenz1})-(\ref{lorenz2})-(\ref{lorenz3}) on the $z_2z_3$-plane.}
\label{fig4}
\end{figure}

We again set $\sigma=10$, $r=28$, $b=8/3$ in system (\ref{lorenz1}), and $\overline \sigma=10$, $\overline r=12$, $\overline b=8/3$, $\mu_1=\mu_2=\mu_3=0.1$, $f_1(x_1,x_2,x_3)=7.3x_1+\cos x_1$, $f_2(x_1,x_2,x_3)=1.2 x_2+0.5 \arctan x_2$, $f_3(x_1,x_2,x_3)= 3.5 x_3+0.9 e^{-x_3}$ in system (\ref{lorenz2}) as in Section \ref{sec3}. Figure \ref{fig3} shows the projections of four chaotic trajectories of the $9$-dimensional system (\ref{lorenz1})-(\ref{lorenz2})-(\ref{lorenz3}) on the $z_1z_2z_3$-space. Moreover, we represent in Figure \ref{fig4} the projections of the same chaotic trajectories on the $z_2z_3$-plane. The color of each of the trajectories depicted in Figure \ref{fig3} is the same with the color of its counterpart shown in Figure \ref{fig4}. The initial data of these trajectories are provided in Table \ref{table1}. The simulation results shown in Figures \ref{fig3} and \ref{fig4} reveal that four chaotic attractors coexist in the dynamics of the $9$-dimensional system (\ref{lorenz1})-(\ref{lorenz2})-(\ref{lorenz3}).

 \begin{table}[ht]
 \caption{Initial data of the trajectories represented in Figures \ref{fig3} and \ref{fig4}.}  
\centering  
   \begin{tabular}{c c c c c c c c c c} 
   \hline
  & $x_1(0)$ & $x_2(0)$ & $x_3(0)$ & $y_1(0)$ & $y_2(0)$ & $y_3(0)$ & $z_1(0)$ & $z_2(0)$ & $z_3(0)$  \\  \hline
 Trajectory in blue \ \ & $-4.874$ & $-0.221$ & $29.059$ & $-4.812$ & $-3.892$ & $11.912$ & $-2.594$ & $-2.461$ & $3.136$ \\  
  Trajectory in  red \ \ & $1.436$ & $2.501$ & $16.269$ & $4.278$ & $4.565$ & $10.058$ & $2.579$ & $2.484$ & $3.065$ \\  
  Trajectory in  green \ \  & $4.969$ & $6.963$ & $18.602$ & $4.838$  & $3.498$ & $12.779$ & $-1.853$ & $-1.961$ & $2.008$ \\  
   Trajectory in black \ \ & $4.537$ & $5.098$ & $21.273$ & $-4.409$ & $-4.884$ & $10.654$ & $1.949$ & $2.034$ & $2.024$ \\  \hline
  \end{tabular} 
  \label{table1} 
  \end{table}

\section{Conclusions} \label{sec5}
 
In this study we demonstrate under certain conditions that two chaotic attractors coexist in the dynamics of unidirectionally coupled Lorenz systems. High dimensional systems with multiple chaotic attractors can be obtained by applying the same type of coupling provided in Section \ref{sec2} to chains of Lorenz systems, and an example of a $9$-dimensional system possessing four chaotic attractors is revealed in Section \ref{sec4}. This is the first time in the literature that the coexistence of four chaotic attractors is obtained.

Global unpredictable behavior of the weather dynamics is one of the subjects associated with our results. An effort was made in study \cite{Akh15} to answer the question \textit{why the weather is unpredictable at each point of the atmosphere} on the basis of Lorenz systems. The whole atmosphere of the Earth was assumed to be partitioned in a finite number of subregions such that the dynamics of each of them is governed by a Lorenz system with certain coefficients. Considering sensitivity as unpredictability of weather in the meteorological sense, it was further assumed in \cite{Akh15} that there are subregions whose corresponding Lorenz systems admit chaos with the main ingredient as sensitivity and subregions whose corresponding Lorenz systems are non-chaotic. The cases of one stable equilibrium point and stable limit cycle were taken into account for the non-chaotic ones. It was deduced that if a subregion with a chaotic dynamics influences another one with a non-chaotic dynamics, then the latter also becomes unpredictable. 

One can confirm that the results of this paper are complementary to the discussions mentioned in \cite{Akh15} such that if the corresponding Lorenz system of a subregion with a non-chaotic dynamics possesses two stable equilibrium points, then that subregion will become unpredictable under the influence of a chaotic subregion and two chaotic attractors may take place in the dynamics. Moreover,  considering further interactions between chaotic and non-chaotic subregions, multiple chaotic attractors may occur depending on the number of stable equilibrium points of the Lorenz systems corresponding to non-chaotic subregions. These inferences may be helpful for analyzes on the complex dynamics of the atmosphere. Our results may also be used as tools in secure communication \cite{Cuomo93} and  for designing coupled Lorenz lasers \cite{Haken75,Lawandy87} possessing multiple chaotic attractors.

\end{document}